\documentclass[12pt]{article}

\begin{document}


\title{\vskip1cm
Ultralocality on the lattice
\vskip.5cm}
\author{Rafael G. Campos$^*$ and Eduardo S. Tututi$^\dagger$\\ 
Escuela de Ciencias F\'{\i}sico--Matem\'aticas, \\
Universidad Michoacana \\
58060 Morelia, Michoac\'an, M\'exico\\
$^*$E-mail: rcampos@zeus.umich.mx\\
$^\dagger$E-mail:  tututi@zeus.umich.mx\\
Fax number: 52 443 316 6969
}
\date{}
\maketitle
{
\vskip.3cm
\noindent PACS: 11.15.Ha, 11.15.Tk, 02.60.Jh\\
}\\[4cm]

\begin{center} Abstract \end{center}
It is shown that the nonlocal Dirac operator yielded by a lattice model that 
preserves chiral symmetry and uniqueness of fields, approaches to an 
ultralocal and invariant under translations operator when the size of the lattice 
tends to zero.
\vfill
\newpage

The Nielsen-Ninomiya theorem \cite{Nie81} has been challenging the construction 
of quiral field theories on the lattice during two decades. The use of Dirac operators 
$D$ satisfying the Ginsparg-Wilson relation 
\cite{Gin82}
\[
\gamma_5 D+D\gamma_5=aD\gamma_5 D,
\]
is a prospective approach to get chiral fermions with no doublers [chiral symmetry 
remains broken by a term $O(a)$] that avoids a contradiction with the 
Nielsen-Ninomiya theorem and yields an exact symmetry of the fermion action \cite{Lus98}. 
However, these operators cannot be ultralocal but may be local \cite{Hor98} (when the 
interaction range of the operator is bounded by some finite lattice distance the operator is 
termed ultralocal, if it is exponentially bounded, the operator is termed local).
\\
In spite that the locality of the Dirac operator is in general a desired property, nonlocal 
operators have been used to circumvent the restrictions of the Nielsen-Ninomiya theorem 
(see for example \cite{Bic00} and references there in) and they may have 
zero modes satisfying the Atiyah-Singer index theorem \cite{Tin01}. This fact  
strengthen the use of nonlocal operators as an approach to follow in lattice QCD.\\
Recently, a lattice scheme where the fermion doubling and the chiral symmetry breaking 
are absent has been presented in \cite{CamTut1}-\cite{CamTut2}. Such a scheme yields a 
nonlocal Dirac operator, but due to some properties of the differentiation 
matrices used to construct it, the quantum algebras and some operational relations maintain 
their canonical forms on the lattice when the number of nodes tends to infinity (or the size of
the lattice tends to zero) \cite{Cam1}-\cite{Cam2}. The aim of this letter is to show that such 
nonlocal differentiation matrices and therefore the Dirac operator constructed with them, 
approach ultralocal operators (the nonlocality tends to disappear) when the size of the lattice 
tends to zero. This fact represents another building stone in our lattice model since the 
chiral symmetry is exactly preserved with no doublers and (asymptotically) gauge invariance. 
\\
\noindent
{\sl Differentiation matrices.} These matrices, arising naturally in the linear interpolation 
problem of functions, yield exact values for the derivative of polynomial functions at 
certain points selected as lattice sites. Due to this, they have been used to solve some 
boundary value problems (see \cite{Cam1}-\cite{Cam2} and references there in).
The class of functions on which the differential operator to be discretized acts on, 
determines the kind of differentiation matrix to be used. Thus, to get a discrete form of 
an operator acting on functions that fall off rapidly to zero at large distances, we can 
use the skew-symmetric differentiation matrix \cite{CamTut2}
\begin{equation}
(D_x)_{jk}=\cases{0,&{$i=j$},\cr\noalign{\vskip .5truecm}
\displaystyle {{(-1)^{j+k}}\over{x_j-x_k}}, &{$i\not=j$},\cr} 
\label{mdx}
\end{equation}
constructed with the $N$ zeros $x_j$ of the Hermite polynomial $H_N(x)$ as lattice 
points\footnote{Equation (\ref{mdx}) gives a simplified expression for $D_x$.},
whereas to get a discrete form of an operator acting on periodic functions \cite{Cam2}, the 
differentiation matrix
\begin{equation}
(D_t)_{jk}=\cases{0,&{$j=k$},\cr\noalign{\vskip .5truecm}
\displaystyle {(-1)^{j+k}\over{2\sin{{(t_j-t_k)}\over 2} }}, &{$j\not=k$},\cr}
\label{mdt}
\end{equation}
should be used. In this case the lattice points are chosen to be the $N$ equidistant 
points\footnote{The same result is obtained for the set of points $t_j={\pi\over N} (2j-N-1)$.}
\begin{equation}
t_j=-\pi+2\pi j/ N, \quad j=1,2,\cdots,N.
\label{tlat}
\end{equation}
\\
Since the fields are expected to fall off rapidly to zero at large distances 
or one may assume periodic boundary conditions on the fields imposed by
the lattice, we can use differentiation matrices of the type (\ref{mdx}) or 
(\ref{mdt}), or a combination of them, to construct a 
discretized formulation of classical field theory in $3+1$ variables where the fields 
are unique (no doublers), the chiral symmetry is not broken when $m=0$ and the 
gauge invariance is recovered in the limit $N\to\infty$. Some of these properties have 
been shown previously \cite{CamTut1}-\cite{CamTut2} for the differentiation matrix 
(\ref{mdx}).  \\
In this letter we analyze the ultralocality property of these differentiation schemes 
used to construct the Dirac operator. In addition, we show the absence of the fermion 
doubling in the formulation with derivative of the type (\ref{mdt}). 
\\
First of all, let us show the relationship between the differentiation matrices 
$D_x$ and $D_t$. 
Let $j$ and $k$ be a pair of distinct and fixed integers. Since the $n$th zero 
of the Hermite polynomial $H_N(x)$ can be approximated by 
\begin{equation}
x_n={{(2n-N-1)\pi}\over{2\sqrt{2N}}}
\label{xne}
\end{equation}
when $n$ is fixed and $N$ is large enough, we have that
\[
(D_x)_{jk}=2\sqrt{2/N}(D_t)_{jk}=(-1)^{j+k}={\sqrt{2N}\over\pi}{{(-1)^{j+k}}\over{j-k}}.
\]
This relation does not become a matrix equation between $D_x$ and $D_t$ since 
(\ref{xne}) is valid only for fixed $n$ when $N\to\infty$ and therefore, $\sin(t_j-t_k)/2$ 
can not be approximated by $(t_j-t_k)/2$ for elements close to the border of $D_t$. 
Thus, if $N$ is large enough,
\begin{equation}
D_x=2\sqrt{2/N}D_t+O, \label{dxdtm}
\end{equation}
where $O$ is a border matrix whose nonzero elements depend on $N$.
By a $N\times N$ border matrix we mean a matrix $M$ whose elements $M_{jk}$ lying
between $n< j,k\le N-n$ are all of them zero for certain $n<<N$.
\\
Clearly, $D_t$ is a Toeplitz matrix \cite{Hor91} 
and fulfills the requirements of translational invariance on the periodic lattice (\ref{tlat}), 
whereas $D_x$ becomes a Toeplitz matrix 
(except for a border matrix) only for $N$ large enough. Both of them are nonlocal matrices for 
a finite $N$ since $\vert\csc(x)\vert >1/\vert x\vert > \exp( -\vert x\vert)$ in the corresponding 
ranges. However, we can use the Toeplitz property to show that both $D_x$ and $D_t$ approach 
ultralocal matrices almost everywhere when $N\to\infty$. By {\sl almost everywhere}  we mean 
{\sl except for border matrices}.  Such border matrices circumvent the no-go theorem given 
in  \cite{Hor98} and allow us to have an ultralocal Dirac operator almost everywhere with the 
properties given above. To show this, we will use the matrix $D_t$ which has a stronger 
nonlocality. \\[1cm]
{\sl Ultralocality of} $D_t$.
Since matrix $D_t$ is a skew-symmetric Toeplitz matrix, it can be written as
\begin{equation}
D_t=\sum_{n=1}^{N-1}d_n(A^n-B^n), \label{dtab}
\end{equation}
where 
\begin{equation}
d_n={(-1)^{n+1}\over{2\sin(n\pi/N)}}
\label{dn}
\end{equation}
and $A$ and $B$ are the backward shift and forward shift matrices respectively, i.e, 
\[
A_{jk}=\delta_{j,k-1},\qquad  B_{jk}=\delta_{j-1,k}.
\]
These matrices satisfy the useful relations
\begin{equation}
ABA=A,\qquad BAB=B, 
\label{relAB}
\end{equation}
which  can be used to show that 
\[
A^{2m}-B^{2m}=(A-B)\sum_{n=0}^{m-1}(A^{2n+1}+B^{2n+1})-
[A,B]\Bigl(I+\sum_{n=1}^{m-1}(A^{2n}+B^{2n})\Bigr)
\]
and
\[
A^{2m+1}-B^{2m+1}=(A-B)\Bigl(I+\sum_{n=1}^{m}(A^{2n}+B^{2n})\Bigr)-
[A,B]\sum_{n=0}^{m-1}(A^{2n+1}+B^{2n+1}).
\]
Here, $I$ stands for the $N\times N$ identity matrix. Thus, (\ref{dtab}) can be rewritten 
in the useful form 
\begin{equation}
D_t=(A-B)\Delta - [A,B]\Sigma, 
\label{dtabmb}
\end{equation}
where $\Delta$ and $\Sigma$ are the symmetric matrices
\begin{equation}
\Delta= s_1 I + \sum_{n=1}^{N-2}s_{n+1}(A^{n}+B^{n}), \quad
\Sigma=s_2 I + \sum_{n=1}^{N-3}s_{n+2}(A^{n}+B^{n}),
\label{mneq}
\end{equation}
and 
\begin{equation}
s_{2m+1}=\sum_{n=m}^{[N/2]-1}d_{2n+1}, \qquad s_{2m}=\sum_{n=m}^{[(N-1)/2]}d_{2n}.
\label{sisp}
\end{equation}
Here, $[x]$ stands for the greatest integer less or equal to $x$. Additionally to (\ref{relAB}), 
$A$ and $B$ satisfy 
\[
([A,B])_{jk}=\cases{1,& $j=k=1$,\cr -1,&$j=k=N$,\cr 0,& otherwise.}
\]
Therefore, the product 
\[
O_1=[A,B]\Sigma
\]
appearing in (\ref{dtabmb}) becomes a border matrix. For $N$ great enough, the sums 
(\ref{sisp}) appearing in (\ref{mneq}) can be approximated by
\[
s_n=(-1)^{n+1}{N\over{4\pi}}\displaystyle{\log \Biggl[{{N\epsilon_n}\over \pi}
\cot({{n\pi}\over{2N}})\Biggl]}, \quad n=1,2,\ldots ,N-1,
\]
where 
\[
\epsilon_n=\cases{1,&$N+n$ even,\cr 2,& $N+n$ odd.}
\]
Since $D_t=(A-B)\Delta$ (except for the border matrix $O_1$), the interaction range of 
$D_t$ is essentially the one of $\Delta$, which is measured by $\vert s_{\vert j-k+1\vert }\vert$.
If we put $x=n\pi/N$, $\vert s_n\vert$ can be rewritten as
\[ 
\vert s_n\vert \equiv \vert s(x)\vert={N\over{4\pi}}
\displaystyle{ \Bigg\vert \log \bigg\vert {N \epsilon_n \over \pi}\cot(x/2)\bigg\vert \Bigg\vert},
\quad x\in(0,\pi).
\]
Note that $s(x)$ is an antisymmetric function in $(0,\pi)$ with respect to $\pi /2$ and that
the absolute values of $s_1\equiv s(0)$ and $s_{N-1}\equiv s(\pi)$ grow as 
$(N/4\pi)\log(2\epsilon_n N^2/\pi^2)$. Thus, for $x$ other than $0$ or $\pi$,
\begin{equation}
{{s(x)}\over{s(0)}}\to 0,\quad N\to\infty. 
\label{delu}
\end{equation}
This is the behavior of an unnormalized Dirac's delta. Thus, $s(x)\to C_N[\delta(x)+\delta(x-\pi)]$.
To get the constant of normalization the measure $\delta x=\pi /N$ has to be taken into account. 
For convenience, we integrate from $-\pi /2$ to $\pi/2$ to obtain
\begin{equation}
\Big[{N^2\over{4\pi}}\log({{N\epsilon_n}\over{\pi}})\Big]^{-1}\int_{-\pi/2}^{\pi/2} s(x) dx \to 1.
\label{deld}
\end{equation} 
Eqs. (\ref{delu}) and (\ref{deld}) show that the asymptotic behavior of $s(x)$, $x\in (-\pi/2,\pi/2)$, 
is that of the Dirac delta multiplied by a growing numerical factor. Therefore, except for the measure,
\begin{equation}
\vert \Delta(x-y)\vert \to {N\over{4\pi}}\log({{N\epsilon_n}\over{\pi}})
\Big[\delta(x-y)+\delta(\vert x-y\vert-\pi)\Big], \quad N\to\infty.
\label{delxy}
\end{equation}
Since $s(\pi)$ corresponds to the elements of $\Delta$ located on the band defined by 
$s_{N-1}$, the second delta of the right-hand of (\ref{delxy}) yields another border matrix. 
Let us denote by $O_2$ such a matrix. Then, we have from (\ref{dtabmb}) that $D_t$ 
approaches an ultralocal matrix as $N\to\infty$ except for the sum of the border matrices
$O_1$ and $O_2$. If $D_t$ is applied to a vector formed with the values of a function 
$f(x)$, the matrix-vector multiplication changes into an integration in the asymptotic limit 
whenever the measure is taken into account. Therefore, the action of 
$D_t$ on such a vector is mainly given by
\begin{equation}
\sum_y D_t(x-y)f(y)\to [f(x+a)-f(x-a)]/2a, \quad a\to 0,
\label{dact}
\end{equation}
where 
\[
a^{-1}={{N^2}\over{4\pi}}\log({{N\epsilon_n}\over{\pi}}).
\]
This is an expected result since $D_t$ is an exact differentiation matrix for certain 
subclasses of periodic functions \cite{Cam2}. Due to (\ref{dxdtm}), the differentiation 
matrix $D_x$ also satisfies a relation like (\ref{dact}).
\\[1cm]
{\sl Uniqueness of the free fermionic fields}. Next, we show that the use of the 
differentiation matrix (\ref{mdt}) avoids the doubling fermion problem on a
lattice. Let us consider a four-dimensional regular lattice with $N^4$ points whose
support are in the interval $-L< x_n^{\mu}\le L$ [substitute $\pi$ by $L$ in (\ref{tlat})].
We assume functional periodicity in the fermion fields as trigonometric polynomials. 
The (continuum) Euclidean free Lagrangian  reads  
\[
{\cal L}=\overline{\psi}(x)\left(\gamma_{\mu}\partial_{\mu}+m\right){\psi}(x)
\, .
\]
We make the substitution 
\begin{equation}
\partial_{\mu}{\psi}(x)\to \left[D_{\mu}\psi\right]_n =
\sum_{l_{\mu}=1}^{N-n_{\mu}}
d_{l_{\mu}}\psi_{n+l_{\mu}}- 
\sum_{l_{\mu}=1}^{n_{\mu}-1}d_{l_{\mu}}\psi_{n-l_{\mu}}
\, ,
\label{derivative}
\end{equation}
where we have used (\ref{dtab}) and (\ref{dn}), and $\mu=1,2,3,4$. Thus 
the discretized action is
\[
S=\sum_{n,\mu} \left[\sum_{l_{\mu}=1}^{N-n_{\mu}}
d_{l_{\mu}}\overline{\psi}_n\gamma_{\mu}\psi_{n+l_{\mu}}- 
\sum_{l_{\mu}=1}^{n_{\mu}-1}d_{l_{\mu}}
\overline{\psi}_n\gamma_{\mu}\psi_{n-l_{\mu}} +m\overline{\psi}_n\psi_n
\right].
\]
At first glance this is a rather nonlocal action, since the field at each lattice 
point couples with the  field evaluated at the reminder points  along each axis. 
Nevertheless, due to the fact that $[D_{\mu} \psi]_n$ is the exact partial 
derivative along the direction $\mu$ at the point $n$ (see \cite{Cam2}), 
the action can be easily diagonalized by taking the discrete Fourier 
transform in the field
\[
\psi_n ={1\over \sqrt{N^4}}\sum_{k}\tilde{\psi}_{k}e^{ik\cdot x_n},
\]
and in its derivative 
\begin{equation}
[D_{\mu} \psi]_n={1\over \sqrt{N^4}}\sum_{k}ik_{\mu}
\tilde{\psi}_{k}e^{ik\cdot x_n}\, ,
\label{ft1}
\end{equation}
here $k_{\mu}=\pi l_{\mu}/L$, with $l_{\mu}$ being an integer restricted to 
values in the range $-(N-1)/ 2< l_{\mu}\leq (N-1)/ 2$ ($N$ is assumed to be 
an odd number). Thus the maximum
momentum allowed is $ |k_{\rm max}|=\pi/ 2\Delta$, with $\Delta$
the length between two successive points. It is clear that the values of $k$
become to be  with no restriction in the limit when $N$ goes to infinity and 
sum in
(\ref{ft1}) must be replaced by an integral in the corresponding Brillouin 
zone. In fact, in this limit the 
derivative in Eq. (\ref{derivative}) tends to the exact continuum value, and 
the continuum theory is recovered.  The Fourier transform of the action 
can be written as
\[
S=\sum_{k,\mu} \overline{\tilde{\psi}}_k\left(i\gamma_{\mu}k_{\mu}+m\right)
\tilde{\psi}_k 
\, ,
\]
which yields to the dispersion relation
\begin{equation}
S(k)=\gamma_{\mu}k_{\mu} +m
\label{disrel}
\, .
\end{equation}
It is clear that no doublers appear since the values of $k$ are proportional
to an integer. Moreover in the limit in which the spacing $\Delta$ tends to
zero (\ref{disrel}) gives the exact energy values. 
\\
Other desired properties, as gauge invariance and chiral symmetry, also 
have support on this scheme. They can be shown as in Ref. \cite{CamTut2}.
Let us emphasize that the exact chiral symmetry is preserved explicitly in 
our scheme as it happens with the so-called SLAC derivative \cite{Dre76}, 
which is {\sl defined} through an equation like (\ref{ft1}). In that context, 
matrix (\ref{mdt}) is an explicit realization of such a derivative with a number
of properties given in \cite{Cam2}.

\end{document}